\documentclass[conference]{IEEEtran}
\IEEEoverridecommandlockouts    
\usepackage{subcaption}
\usepackage{fourier}

\usepackage{microtype}
\usepackage{cite}
\usepackage{amsmath,amssymb,amsfonts}
\usepackage{algorithmic}
\usepackage{graphicx}
\usepackage{multirow}
\usepackage{textcomp}
\usepackage{xcolor}
\usepackage{float}
\usepackage[numbers,sort&compress]{natbib}
\usepackage[colorlinks=true,linkcolor=blue,citecolor=blue,urlcolor=blue]{hyperref}
\def\BibTeX{{\rm B\kern-.05em{\sc i\kern-.025em b}\kern-.08em
    T\kern-.1667em\lower.7ex\hbox{E}\kern-.125emX}}

\newcommand{\Remark}[1]{{\color{red}}}

\definecolor{todocolor}{RGB}{255, 20, 147}  
\definecolor{todoBg}{RGB}{255, 240, 245}    

\newenvironment{multiequation}{%
  \setlength{\abovedisplayskip}{3pt}      
  \setlength{\belowdisplayskip}{4pt}      
  \setlength{\abovedisplayshortskip}{3pt} 
  \setlength{\belowdisplayshortskip}{4pt} 
  \begin{equation}
    \begin{aligned}
}{%
    \end{aligned}
  \end{equation}
}

\usepackage{caption}
\title{\LARGE \bf Car-following Models and Congestion Control with Followerstopper on a Ring-Road under Known Delay -- Examining Limit Cycle}

\author{
	\parbox{\textwidth}{%
		\centering
		Trevor McClain$^{1}$, Rahul Bhadani$^{1}$
	}%
	\thanks{$^{1}$AI, Autonomy, Resilience, Control Lab (AARC Lab) Department of Electrical \& Computer Engineering The University of Alabama in Huntsville, Huntsville, USA 
		{\tt\small thm0013@uah.edu, rahul.bhadani@uah.edu}}%
}

\begin{document}
	
	\maketitle
	\thispagestyle{empty}
	\pagestyle{empty}
	
	\begin{abstract}
		  This paper examines the IDM microscopic car-following model from a dynamical systems perspective, analyzing the effects of delay on congestion formation. Further, a case of mixed-autonomy is considered by controlling one car with Followerstopper in a ring road setting containing IDM vehicles as human drivers. Specifically, the stop-and-go waves phenomenon in idealized traffic from a dynamical systems perspective is examined. We show that Followerstopper-controlled vehicle is effective at eliminating emergent stop-and-go waves in the IDM traffic simulation. We show through simulation that the uniform flow manifold is unstable for the ring road simulation with IDM vehicles, and that replacing a single car with Followerstopper induces stability, allowing the cars to drive safely at a uniform speed. Additionally, the case of known delay is considered in a mixed-autonomy scenario. Our simulation result shows that while considering a known time delay, traffic waves emerge earlier than in the no-delay case. At the same time, a single-vehicle controlled using Followerstopper controller is able to prevent the emergence of traffic waves even in the presence of delay.
          \end{abstract}
	
	\section{Introduction}
	\label{sec:introduction}
    \vspace{-10pt}
	Car-following models are mathematical frameworks for studying the dynamics of how a vehicle follows another. The first such model, the General Motors model \cite{chandler1958traffic}, was introduced nearly 75 years ago. Its successor, the GHR model \cite{gazis1961nonlinear}, and more recent popular models like the IDM \cite{20_Treiber_2000}, have been widely used by researchers since its inception in 2000. Recently, some limitations and improvements have been examined by Albeaik et al. \cite{albeaik2022limitations}, and Zhang et al. In \cite{zhang2024bayesian}, the authors experimented with using different types of Bayesian calibration to optimize the IDM model parameters for realistic driving patterns. From a dynamical systems perspective, multi-vehicle microscopic driving model simulations are many-body dynamical systems. Using this representation, the dynamics of the system can be analyzed for emergent behavior that could not be seen by analyzing a single leader-follower pair. This paper examines the dynamical system created from simulating IDM vehicles on a ring road. We examine the effect of delay on the system, as well as the effects of replacing one of the cars with Followerstopper model on congestion improvement by creating a mixed-autonomy scenario~\cite{stern2018dissipation, bhadani2018dissipation, bhadani2025followerstopper}.

    Early car-following models, such as the General Motors model \cite{doi:10.1287/opre.6.2.165} and its successor the GHR model \cite{gazis1961nonlinear}, operated on a simple principle: the follower vehicle's desired acceleration was proportional to the velocity difference with the leader, multiplied by a sensitivity factor and a time delay. Research at this stage focused primarily on calibrating these parameters against real-world data. Consequently, the goal was not to create a truly autonomous vehicle, but rather to model realistic human driving behavior.

In the 1970-80s, the focus of car-following models shifted towards mimicking not only realistic, but safe driving. With the invention of Gipps' model \cite{gipps1981behavioural} that chooses the lesser of two velocities, one being a safe-traffic-free velocity, and the other being a safe congested-traffic velocity that allows for safe stopping in the case of an abrupt stop from the car in front.

In the 1990-2000s, more complicated models were created, such as the optimal velocity model (OVM)~\cite{lazar2016review} and the intelligent driver model (IDM). The optimal velocity model experimented with different velocity functions to enhance driving performance or increase stability. The IDM model combined terms to represent the desired velocity in free-flow traffic and follow-the-leader style of traffic control, allowing for a single function that represented desired behavior for all conditions.

In the post-2000s era, researchers focus on creating enhanced models with machine learning, physics-informed deep learning \cite{Mo_2021}, and reinforcement learning \cite{khan2025towards}. Additionally, focus has been placed on calibrating older models in real-time to enhance safety and traffic flow. For example, Chengyuan Zhang and Lijun Sun  \cite{zhang2024bayesian} developed a Bayesian calibration method for choosing IDM model parameters in real-time. One of the most important models of the last decade, Followerstopper model \cite{stern2018dissipation}, shifted the focus of autonomous vehicles toward behavior that optimized traffic flow, while maintaining safe driving patterns.

    The remainder of this work is split into the following sections. The background section gives an overview of the IDM, the state space representation of traffic flow, critical points, invariant subspaces, limit cycles, the fundamental diagram of traffic flow, Lyapunov exponents, and a review of Followerstopper. The methodology section describes in detail the experimental setup for the simulations. Finally, the results section describes the relevant results gathered from each of the trials.

        \vspace{-6pt}
        
\section{Background}
   
    \vspace{-5pt}
    
\subsection{Intelligent Driver Model}

    \vspace{-5pt}
    
The Intelligent Driver Model (IDM) \cite{20_Treiber_2000} is a parsimonious car-following model in which a follower vehicle computes a safe acceleration at each time step based on its leader’s position and speed. This acceleration adjusts the follower’s speed to maintain a collision-free headway. For example, if the leader decelerates, the model yields a negative acceleration (Equation~\eqref{eq:IDM}), prompting the follower to slow down accordingly.

{\footnotesize
\begin{equation}
\label{eq:IDM}
\alpha_{\text{IDM}}(s, v, \Delta v) = a \left[1 - \left( \frac{v}{v_0} \right)^\delta - \left( \frac{s^*(v, \Delta v)}{s} \right)^2 \right]
\end{equation}
}

In Equation~\eqref{eq:IDM},  $s^*$ is defined as $
s^*(v, \Delta v) = s_0 + vT + \frac{v \Delta v}{2 \sqrt{ab}}$ where $a$ is the maximum acceleration, $v$ is the current speed, $v_0$ is the desired speed, $\Delta v$ is the speed difference between the leading and following vehicle, $s$ is the space headway between the leading and following vehicle, $s_0$  is the desired standstill distance, $T_0$ is the desired time-gap, and $b$ is the vehicle’s comfortable deceleration rate. 

    \vspace{-5pt}
    
\subsection{Followerstopper Model}

    \vspace{-5pt}
    
Followerstopper model \cite{stern2018dissipation} is relatively unique in its approach to vehicle control, maintaining safety guarantees from previous models while also optimizing for traffic flow. 
The control law for a vehicle  driven using Followerstopper can be written as

\vspace{-10pt}

{\footnotesize
\begin{multiequation}
    \label{eq:Followerstopper}
& v_{\textrm{cmd}}(t) =  f_{\textrm{FS}}(\Delta x(t), \Delta v(t), v_{lead}(t))  = \\
& \begin{cases}
    0,  &\textrm{if} (\Delta x,\Delta v_i) \in \mathcal{S}_1\\
    v(v_{lead}) \frac{\Delta x-d_1(t)}{\bar{d}_2(t)-d_1(t)}, &\textrm{if}  (\Delta x,\Delta v) \in \mathcal{S}_2\\
    v(v_{lead})+(r-v(v_{lead}))\frac{\Delta x-d_2(t)}{d_3(t)-\bar{d}_2(t)}, &\textrm{if} (\Delta x,\Delta v) \in \mathcal{S}_3\\
    r, &\textrm{if} (\Delta x,\Delta v) \in \mathcal{S}_4 
    \end{cases}
\end{multiequation}
}

where $v : \mathbb{R} \rightarrow \mathbb{R}$ is $v(v_{lead}) = \min\{ \max\{v_{lead},0\},r\}$. Four sets $\mathcal{S}_1$, $\mathcal{S}_2$, $\mathcal{S}_3$, and $\mathcal{S}_4$ divided by three safety envelopes as are defined below:

\vspace{-10pt}

{\footnotesize
\begin{multiequation}
     \label{eq:FS_regions}
             \mathcal{S}_1 &= \big\{(\Delta x_i,\Delta v_i) \in \mathbb{R}^2 | 
    0 < \Delta x_i \leq d_1(\Delta v_i) \big\}, \\
    \mathcal{S}_2 &= \big\{(\Delta x_i,\Delta v_i) \in \mathbb{R}^2 | d_1(\Delta v_i) < \Delta x_i \leq d_2(\Delta v_i) \big\}, \\
    \mathcal{S}_3 &= \big\{(\Delta x_i,\Delta v_i) \in \mathbb{R}^2 |d_2(\Delta v_i) < \Delta x_i \leq d_3(\Delta v_i) \big\}, \\
    \mathcal{S}_4 &= \big\{(\Delta x_i,\Delta v_i) \in \mathbb{R}^2 |d_3(\Delta v_i) < \Delta x_i \big\}.
\end{multiequation}
}

and switching boundary $d_j : \mathbb{R} \rightarrow \mathbb{R}$ are:
{\footnotesize
\begin{multiequation}
    d_j(\Delta v_i) = \omega_j + \frac{1}{2\alpha_j} \min \{0,\Delta v_i\}^2, \quad j = 1,2,3,
    \label{eq:FS_envelopes}
\end{multiequation}
\vspace{-10pt}
}

where $\omega_1, \omega_2, \omega_3, \alpha_1 , \alpha_2$, and $\alpha_3$ are controller parameters. $r$ is a free flow desired speed~\cite{bhadani2025followerstopper}.
Followerstopper dissipates stop-and-go waves, maintaining safe driving conditions for all vehicles, by maintaining this variable space headway region, allowing congestion to be absorbed instead of propagated to the next vehicle.

    \vspace{-5pt}
    
    \vspace{-5pt}
    
\subsection{The Dynamical Systems Perspective and State Space Representation of Traffic Flow}

    \vspace{-5pt}
    
A dynamical system describes the evolution of a system through time \cite{lynch2018dynamical}, based on a given update rule and initial conditions. For a traffic flow scenario, with $N$ cars, vectors in the state space $\Sigma$ will be of the following form:

    \vspace{-5pt}
    
{\footnotesize
\begin{equation}
    \label{state_vect}
    z(t) = [ x_1(t),v_1(t), x_2(t),v_2(t), ..., x_N(t),v_N(t),]^T
\end{equation}
}

Additionally, the update rule $\Phi$ can be defined as follows:

    \vspace{-5pt}
    
{\footnotesize
\begin{equation}
    \label{update_rule}
    \dot{z}(t) = f(z(t)) 
\end{equation}
}

In Equation ~\eqref{update_rule}, $f$ represents the car following models used for each vehicle. The dynamical system can also be represented as a system of ordinary differential equations (ODEs), assuming that the $n$-th vehicle is only affected by the vehicle directly in front of it, the ($n$-$1$)-th vehicle:

{\footnotesize
\begin{equation}
    \dot{x}_n(t) = f_x(x_{n-1}(t),v_{n-1}(t),x_n(t),v_n(t)) 
\end{equation}
\begin{equation}
    \dot{v}_n(t) = f_v(x_{n-1}(t),v_{n-1}(t),x_n(t),v_n(t)) 
\end{equation}
}

    \vspace{-5pt}
    
\subsection{Critical points in Dynamical Systems} 
    
    \vspace{-5pt}
    
A critical point in a dynamical system, sometimes called a fixed point, is a point where:

\vspace{-5pt}

{\footnotesize
\begin{equation}
    \dot{z}(t) = f(z(t)) = 0
\end{equation}
}

This implies, that for $t_0$ where $\dot{z}(t_0) = f(z(t_0)) = 0$, then for all $t_{s} > t_0$, $\dot{z}(t_s) = f(z(t_s)) = 0$.
For a dynamical system where the state vector is of size $N$, at each linearizable critical point, a critical point is said to be unstable or stable, depending on the types and quantities of the eigenvalues and eigenvectors \cite{lynch2018dynamical}.
If a critical point $x_0$ is stable, then for a given $\epsilon >0$ there exists $\delta > 0$ where for all $t>t_0$, $||x(t) - x_0(t)||<\epsilon \text{ whenever } ||x(t) - x_0(t)||<\delta $. Put plainly, a critical point is considered stable if there exists a local attraction such that points that start arbitrarily close to the critical point remain arbitrarily close to the critical point. A point is considered unstable if this is not the case. A focus is a critical point, either stable or unstable, where the corresponding eigenvalue pair is complex and the system dynamics form a spiral pattern around the focus. See Figure~\ref{fig:IDM_PS}  (left) for an example of an unstable focus.

    \vspace{-5pt}
    
\subsection{Invariant Subspace and Uniform Flow Manifold in Dynamical Systems}

    \vspace{-5pt}
    
An invariant subspace, $W$, in the context of dynamical systems is a subset of $\Sigma$ where, for a given update rule $\phi()$ and initial trajectory $x_0$, if $x_0 \in W$ then $\phi^t(x_0) \in W, \forall t \ge 0 $.

And, the uniform flow manifold is a special case of an invariant subspace where, in addition to the above, the following condition holds for all trajectories in $W$.
\begin{equation}
    v_n(t) = v^*, \ \forall n
\end{equation}
\begin{equation}
    s_n(t) = s^*, \ \forall n
\end{equation}

These conditions enforce equal spacing and velocity on all vehicles in a platoon, maximizing the flow of traffic, and are often the initial positions of vehicles in simulations or experiments \cite{stern2018dissipation} that seek to quantify the causes and effects of stop-and-go waves.

    \vspace{-5pt}
    
    \subsection{Limit Cycles in Car-Following Models as Dynamical Systems}
        
        \vspace{-5pt}
        
A limit cycle is an isolated periodic solution to a dynamical system \cite{lynch2018dynamical}. A limit cycle represents an oscillation of the system around a center. In physical systems such as traffic flow, this represents a continuous change in relative velocity and space headway between the leader and follower cars. A limit cycle can be stable, unstable, or bi-stable. If a limit cycle is stable, then for a given neighborhood of values in the state space around the limit cycle, these values are attracted to the limit cycle. A limit cycle is said to be unstable if, for a given neighborhood of values in the state space around the limit cycle, these values are repelled from the limit cycle. A limit cycle is bi-stable if there exist two or more neighborhoods around the limit cycle that exhibit stable and unstable dynamics.

    \vspace{-5pt}
    
\subsection{Fundamental Diagrams for Car-Following Models}

    \vspace{-5pt}
    
The fundamental diagram for traffic flow and car-following is defined as a graph that plots the density vs flow of cars on a road or in traffic simulations. The density, $k$, on a road is a measure of how many cars are in a given road length, and is expressed in $\text{cars/meter}$. The flow is a measure of how many cars pass a given area, and is expressed in $\textrm{cars/second}$. Additionally, note that the relationship between velocity, $v_n(t)$, of cars in the area, in $\textrm{meters/second}$, flow, and density is defined as follows. 

\vspace{-10pt}

{\footnotesize
\begin{equation}
    v_n(t) = \frac{q_n(t)}{k_n(t)} \implies q_n(t) = k_n(t) v_n(t)
\end{equation}
}

There are many ways to estimate the flow of agents \cite{makridis2025fundamental}, or a specific agent \cite{zhang2011transitions}, in system simulations. For this paper, velocity is gathered directly from the simulation. Density is estimated with a 1-D Voronoi approximation of density, sometimes called piecewise constant kernel density estimation, defined here as follows.

\vspace{-10pt}

{\footnotesize
\begin{equation}
    k_n(t) = \frac{1}{\Delta x_n(t)}
\end{equation}
}

    \vspace{-5pt}
    
\subsection{Lyapunov Exponents of Dynamical Systems}  
\label{sec:lyapunov}

 \vspace{-5pt}
 
Lyapunov exponents are a technique used in dynamical systems theory to quantify uncertainty~\cite{melin2001modelling}. Specifically, Lyapunov exponents measure the sensitivity to initial conditions. For two initial trajectories $z_1,z_2 \in \Sigma$ with initial separation $\delta_0$, their divergence, or rate of separation, is given by 
$|\delta(t)| \approx e^{\lambda t}|\delta_0|$
Where $\lambda$ is the Lyapunov exponent, the Lyapunov exponent is a measure of the rate of separation or contraction of trajectories as they move through the state space. A positive Lyapunov exponent suggests that the system is chaotic and sensitive to initial conditions, while a negative Lyapunov exponent suggests a lack of sensitivity to initial positions. Each axis in a system will have a different rate of separation, and thus for an $N$-dimensional state space, there will be $N$ Lyapunov exponents. Given the exponential nature of the equation, we care primarily about the maximum Lyapunov exponent, as the effect of smaller Lyapunov exponents will decrease proportionally when compared to the effect of larger Lyapunov exponents.

     \vspace{-5pt}
     
	\section{Simulation Experiment}
	\label{sec:designandimplementation}

 \vspace{-5pt}
 
This study investigates the impact of a uniform time delay on traffic dynamics using a ring-road simulation based on the Intelligent Driver Model (IDM). The analysis is subsequently extended to a mixed-autonomy scenario by replacing one IDM vehicle with a Followerstopper controller to examine its influence on stop-and-go phantom jam formation under known delay conditions.

Four distinct simulation scenarios were configured on a single-lane ring road:
\begin{enumerate}
\item \textbf{Baseline IDM:} 10 IDM vehicles with no delay.
\item \textbf{Delayed IDM:} 10 IDM vehicles with a uniform time delay.
\item \textbf{Mixed Autonomy:} 9 IDM vehicles (no delay) and 1 Followerstopper vehicle.
\item \textbf{Delayed Mixed Autonomy:} 9 IDM vehicles (with delay) and 1 Followerstopper vehicle.
\end{enumerate}

\begin{figure}[H]
    \centering
    \includegraphics[width=0.7\linewidth]{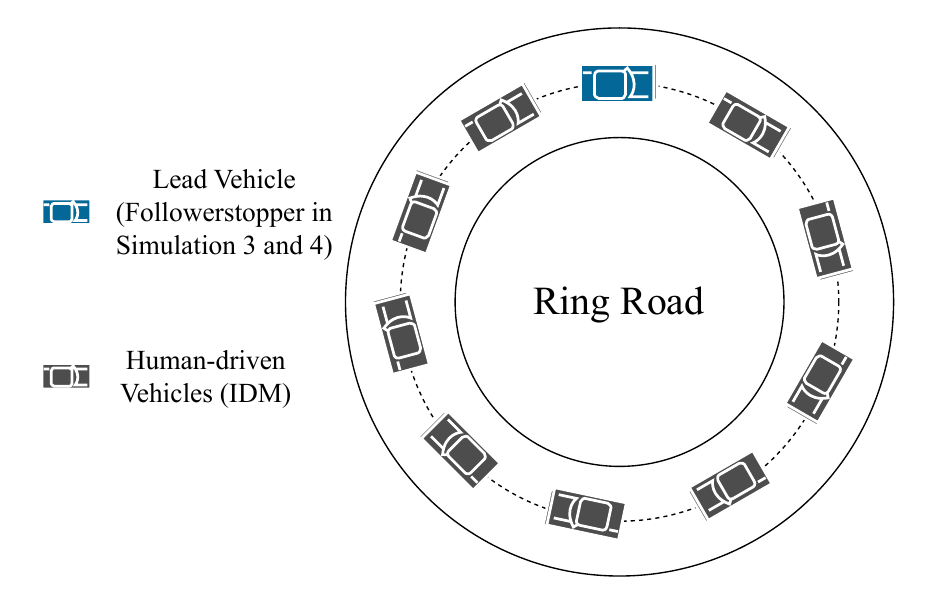}
    \caption{Visualization of the ring-road simulation. The blue car is the leader car that is controlled by an IDM model in simulations $1$ and $2$, and controlled by a Followerstopper model in simulations $3$ and $4$.}
    \label{fig:ring_road_diagram}
     \vspace{-16pt}
     
\end{figure}

We do not consider delay in Followerstopper model as it is an autonomous control, and reaction delay will be negligible compared to reaction delay in human driving. All simulations consist of 10 vehicles. In Scenarios 3 and 4, the lead vehicle (depicted as the blue car in Figure~\ref{fig:ring_road_diagram}) is replaced by Followerstopper controller vehicle. The initial condition for all scenarios is a uniform traffic flow.

  \vspace{-18pt}
  
All vehicles were initialized with uniform spacing on a $100$-meter ring road, consistent with the uniform flow manifold. The initial velocity for each vehicle was set to the uniform flow velocity of $5 \, \text{m/s}$, with a superimposed random perturbation of $1 \times 10^{-3} \, \text{m/s}$. Simulations 1 and 2 (the all-IDM) were run for $1500$ seconds to demonstrate the emergence of stop-and-go waves in the IDM, even in the absence of delay.

In all simulations, the IDM model (Equation $\eqref{eq:IDM}$) had the following parameters: maximum acceleration $a = 0.73$ m/s$^2$, desired speed $v_0 = 33.33$ m/s, desired standstill distance $s_0 = 2$ m, desired time gap $T_0 = 1.6$ s, and comfortable deceleration rate $b = 1.67$ m/s$^2$. These values are considered physically meaningful \cite{albeaik2022limitations}. For Followerstopper controller (Equation $\eqref{eq:Followerstopper}$), the parameters were: desired velocity $r = 4.75$ m/s, initial boundaries $(\omega_{1}, \omega_{2}, \omega_{3}) = (2.25, 3, 4.5)$ m, and deceleration rates $(\alpha_{1}, \alpha_{2}, \alpha_{3}) = (1.0, 0.7, 0.5)$ m/s$^2$.

For Simulations~1 and 2, $\texttt{ODE45()}$ was used to numerically integrate the ODEs. For Simulations~3 and 4, the MATLAB function $\texttt{DDE23()}$ was employed, as $\texttt{ODE45()}$ does not support time delays in differential equation solution.

The Lyapunov exponent was estimated using MATLAB's $\texttt{lyapunovExponent()}$ function. Each solution was linearly interpolated to achieve a uniform sampling frequency of $30$ Hz, satisfying the equal spacing requirement of the $\texttt{lyapunovExponent()}$ function.

 \vspace{-5pt}
 
    \section{Results}

 \vspace{-5pt}

This section analyzes the effects of time delay and the mixed-autonomy scenario with one vehicle controlled by Followerstopper, and human driving is modeled using IDM. For each scenario, the system behavior is visualized through several metrics: the microscopic Voronoi fundamental diagrams, time-space diagrams, phase space diagrams, velocity heatmaps, time-series plots of velocity and space headway.

 \vspace{-5pt}

\subsection{Time-space Diagrams of Vehicle Positions}

\vspace{-5pt}

\begin{figure}[htpb]
    \centering
\includegraphics[width=0.8\linewidth]{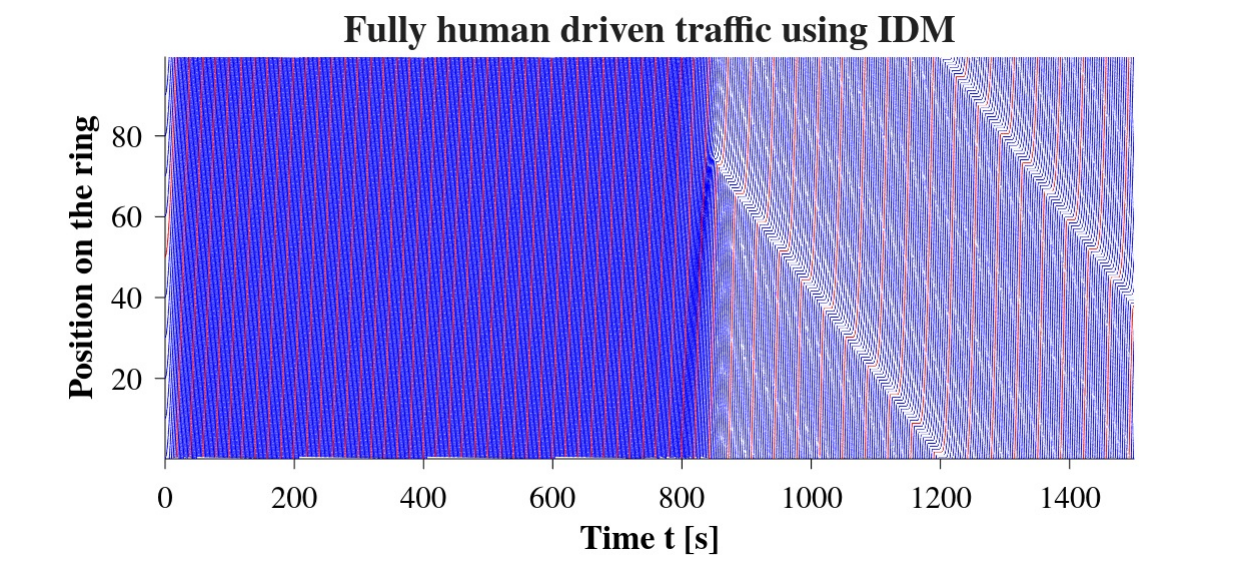}
    \includegraphics[width=0.8\linewidth]{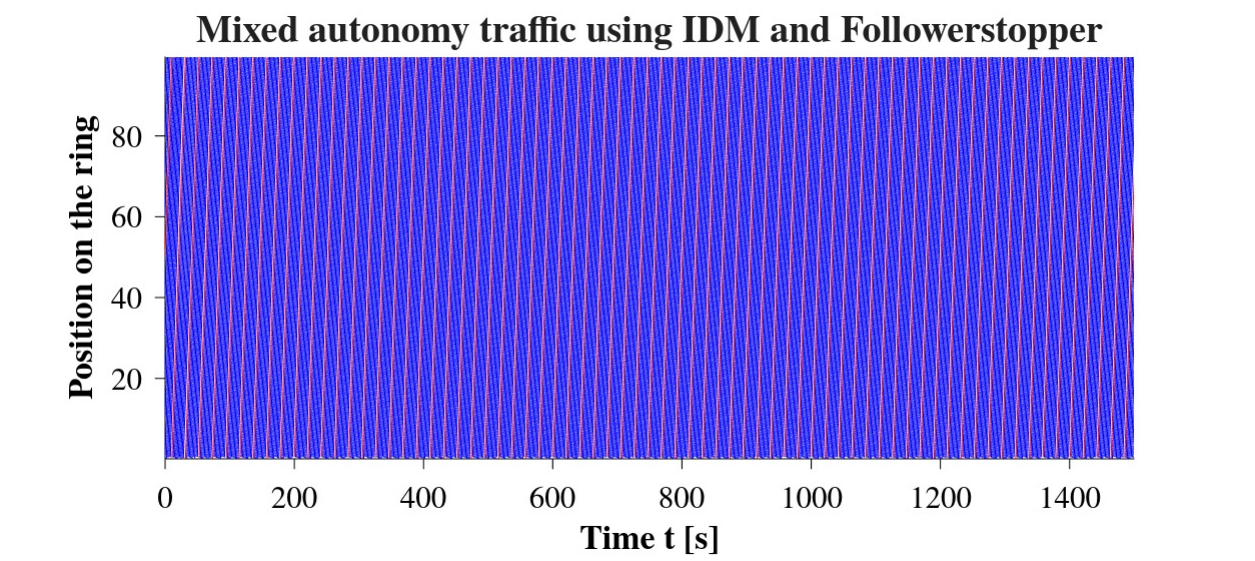}
    \caption{Time-space diagram: \textbf{Top:} IDM simulation; \textbf{Bottom:} Mixed autonomy simulation with the leader vehicle controlled by Followerstopper (red line). No time delay was considered. }
    \label{fig:IDM_pos}

     \vspace{-10pt}
     
\end{figure}

In Figure~\ref{fig:IDM_pos} (top), we can see that the IDM simulation without time delay experienced emergent stop-and-go waves. In contrast, Figure~\ref{fig:IDM_pos} (bottom) shows that in Simulation 3, Followerstopper effectively dissipated the perturbations compared to emergent stop-and-go waves in Simulation 1. 

\begin{figure}[htpb]
    \centering
    \includegraphics[width=0.75\linewidth]{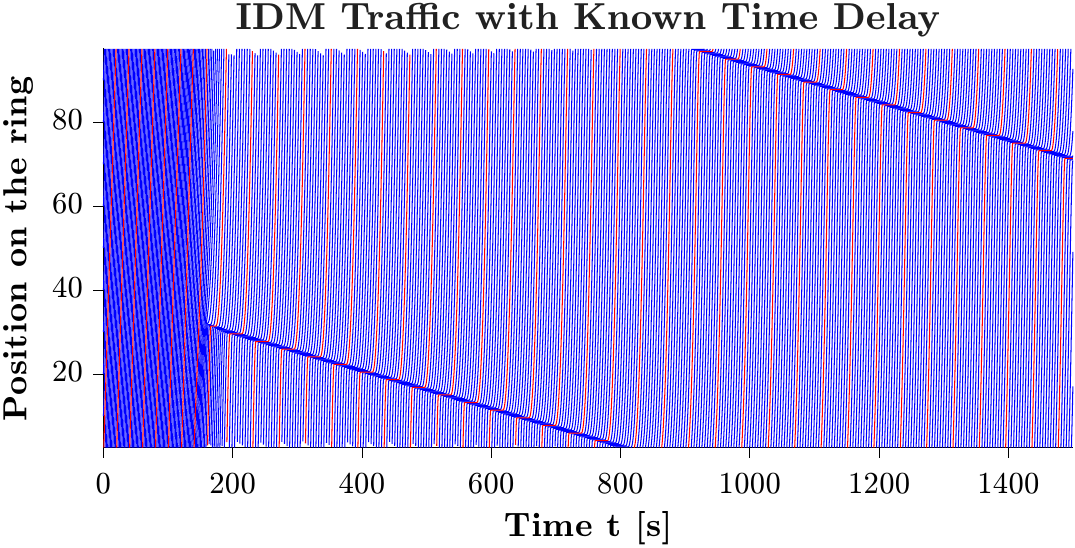}
    
    \includegraphics[width=0.75\linewidth]{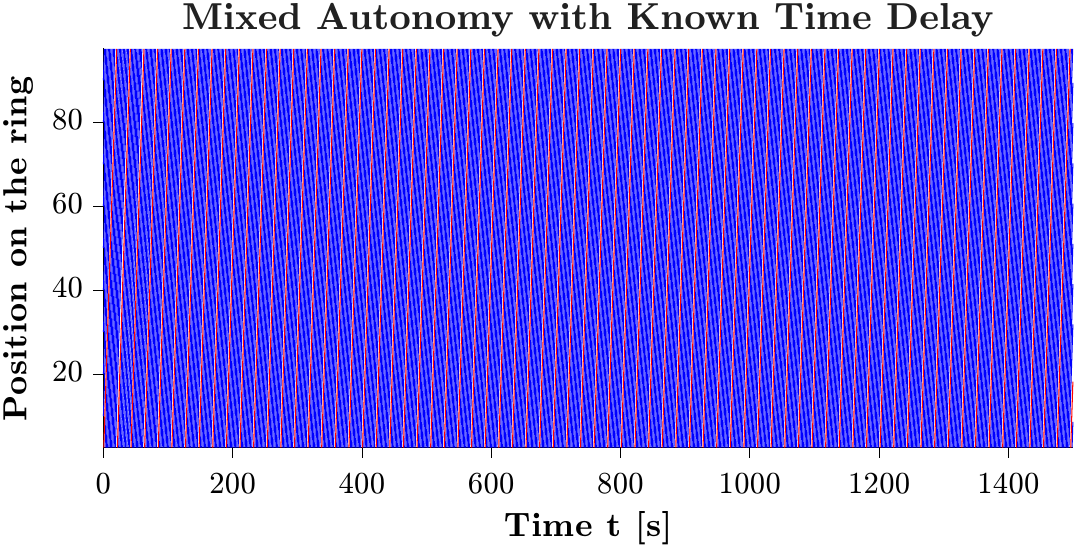}
    \caption{Time-space diagram: \textbf{Top:} delayed IDM simulation; \textbf{Right:} Mixed autonomy simulation with known delay.}
    \label{fig:D_IDM_vpos}

     \vspace{-10pt}
     
\end{figure}

In Figure~\ref{fig:D_IDM_vpos} (top), we can see that considering a time delay of $0.5$ seconds severely affected the IDM simulation, causing stop-and-go waves to appear much sooner, and to be slightly larger than in the zero-delay IDM model. However, no collision occurred, despite the increased size of the stop-and-go waves. In contrast, mixed autonomy simulation consisting of a delayed IDM model and a Followerstopper controlled autonomous vehicle, Figure~\ref{fig:D_IDM_vpos} (bottom), shows close to ideal conditions, maintaining uniform velocity after a small settling period.




\subsection{Fundamental Diagrams}
\begin{figure}[t]
    \centering
    \includegraphics[width=0.45\linewidth]{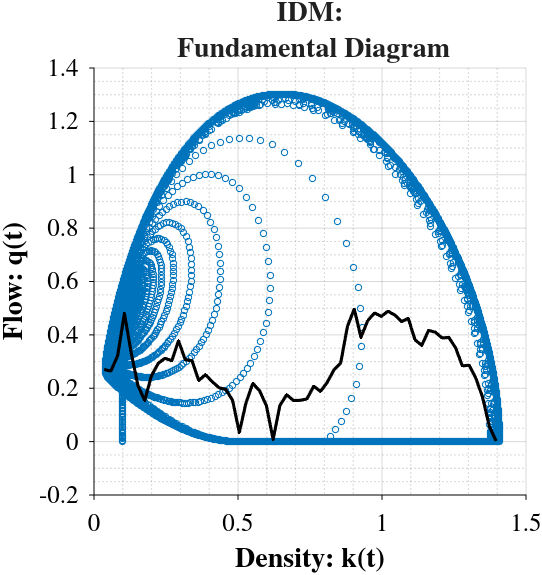}
    \includegraphics[width=0.45\linewidth]{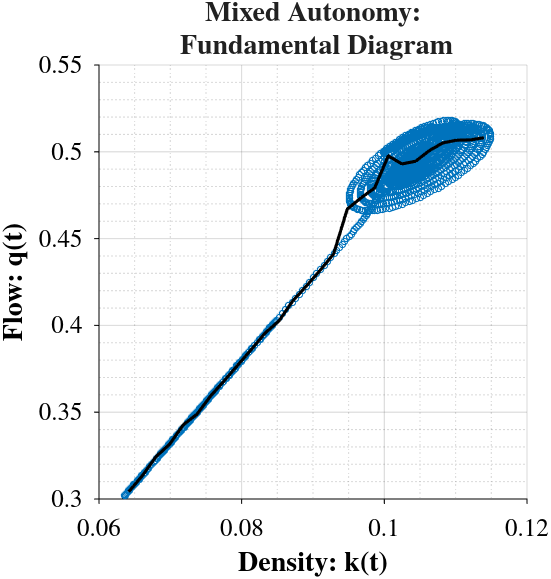}
    \caption{Fundamental diagram using the 1-D Voronoi microscopic approximation in the IDM simulation (left), and Mixed-autonomy simulation (right).}
    \label{fig:IDM_FD}
\end{figure}
\begin{figure}[t]
    \centering
    \includegraphics[width=0.45\linewidth]{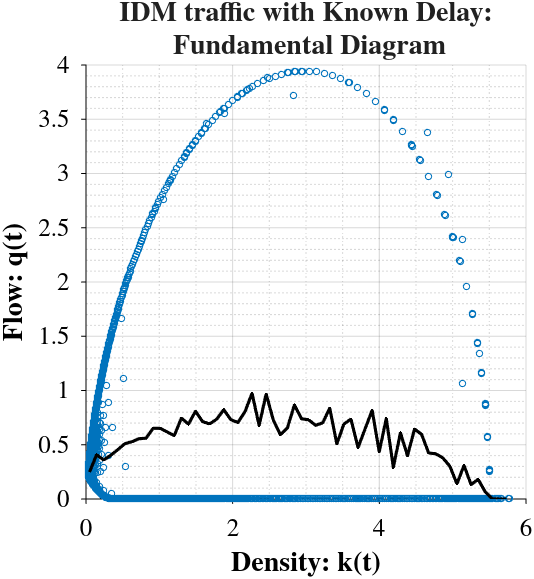}
    \includegraphics[width=0.45\linewidth]{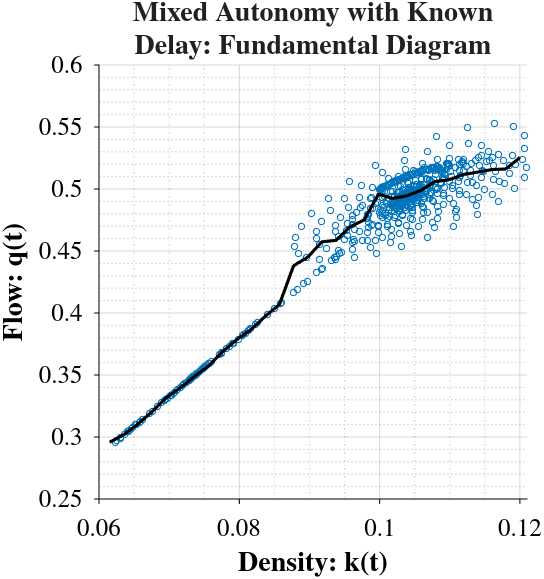}
    \caption{Fundamental diagram using the 1-D Voronoi microscopic approximation in the delayed IDM simulation (left), and delayed IDM with Followerstopper in mixed autonomy simulation (right).}
    \label{fig:D_IDM_FD}
    \vspace{-16pt}
\end{figure}

From Figure~\ref{fig:IDM_FD} (left), we see that the flow generally decreased as density increased. Its maximum density was 1.5 $cars/meter$. This is in contrast with Figure~\ref{fig:IDM_FD} (right), where the simulation mostly maintained a uniform density of $0.1$, equivalent to uniform flow. In the ideal IDM-only simulation, we can see that vehicles came to a complete stop frequently during the final sections of the simulation.

The delayed IDM fundamental diagram, shown in Figure~\ref{fig:D_IDM_FD} (left), shows a much more congested traffic pattern, with cars spending most of their time during the simulation in standstill, high-density traffic, or accelerating to the next stop wave. The peak density in this simulation was also approximately $4$ times as large as that of the ideal zero-delay IDM simulation. In contrast to that simulation, the IDM with Followerstopper simulation shows only a mild increase in density, centering at approximately $0.105$ instead of the ideal $0.1$.

\subsection{Velocity Heatmaps}

\begin{figure}[H]
    \centering
    \includegraphics[width=0.7\linewidth]{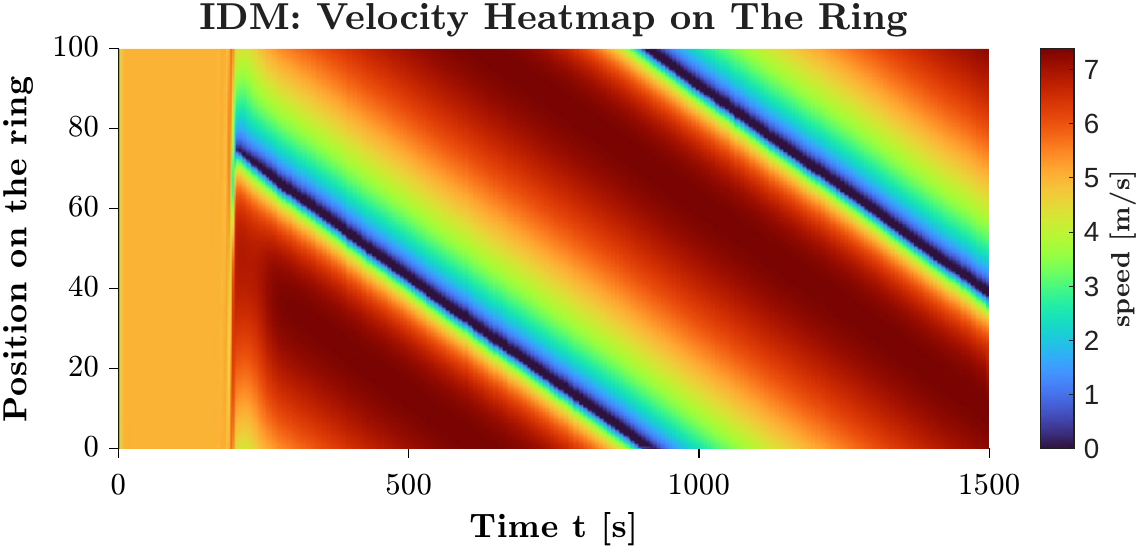}
    \includegraphics[width=0.7\linewidth]
    {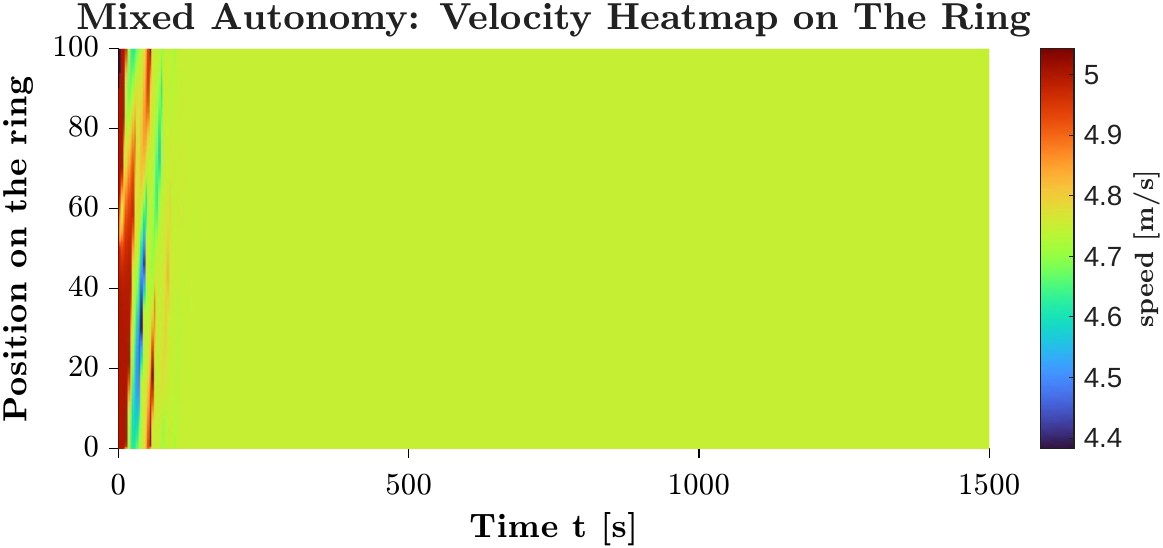}
    \caption{\textbf{Top:} Heatmap showing the velocity on the ring in the IDM simulation; \textbf{Bottom:} Heatmap showing the velocity on the ring in the IDM-Followerstopper mixed autonomy simulation.}
    \label{fig:IDM_vHeat}
\end{figure}

The heatmaps in Figure~\ref{fig:IDM_vHeat} reinforce what is visible in the previous plots, that when perturbations started to magnify, Followerstopper dissipates the perturbations before stop-and-go waves could fully manifest themselves.

\begin{figure}[H]
    \centering
    \includegraphics[width=0.7\linewidth]{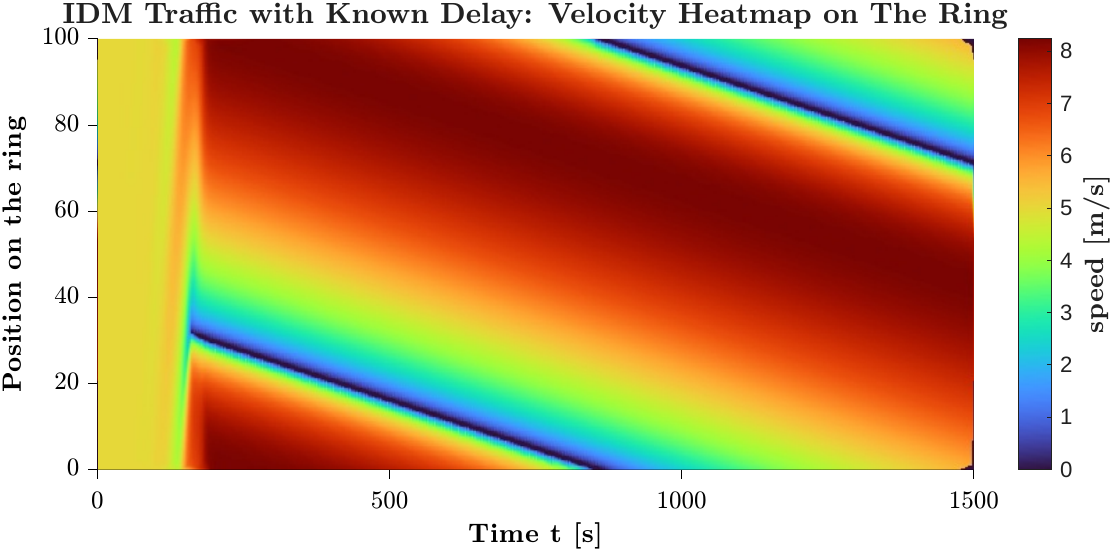}
    \includegraphics[width=0.7\linewidth]{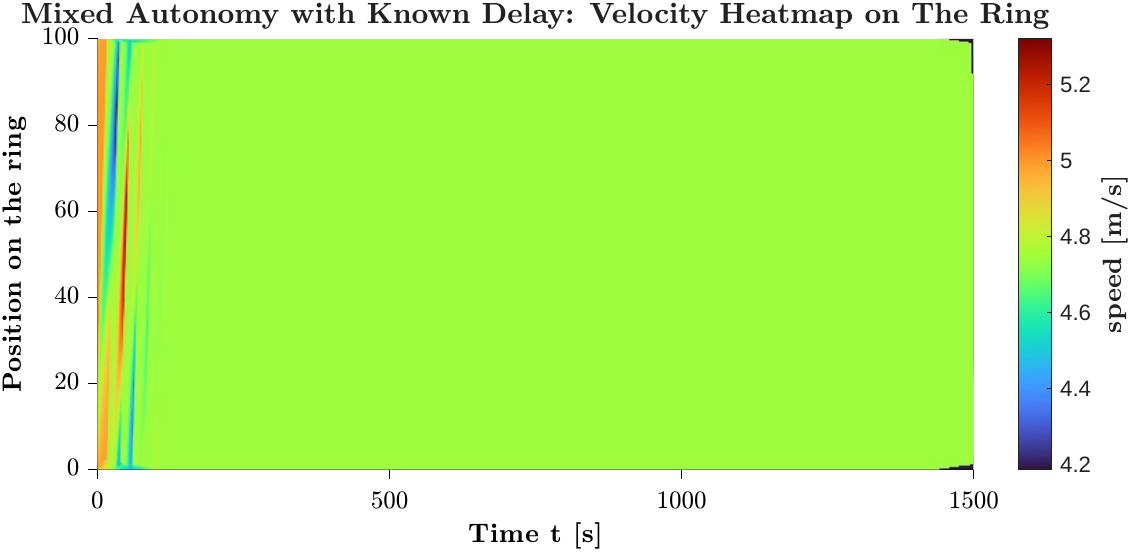}
    \caption{\textbf{Top:} Heatmap showing the velocity on the ring in the delayed IDM simulation; \textbf{Bottom:} Heatmap showing the velocity on the ring in the delayed IDM and Followerstopper mixed autonomy simulation.}
    \label{fig:D_IDM_vHeat}
\end{figure}
From the velocity heatmaps in Figure~\ref{fig:D_IDM_vHeat} (top), we can see that a single stop-and-go wave formed relatively early in the simulation, moving slowly on the ring road in the simulation with time. In contrast, even with a time delay, Followerstopper simulation experiences no significant stop-and-go waves.

\subsection{Velocity and Space Headway Over Time}

\begin{figure}[H]
    \centering
    \includegraphics[width=0.49\linewidth]{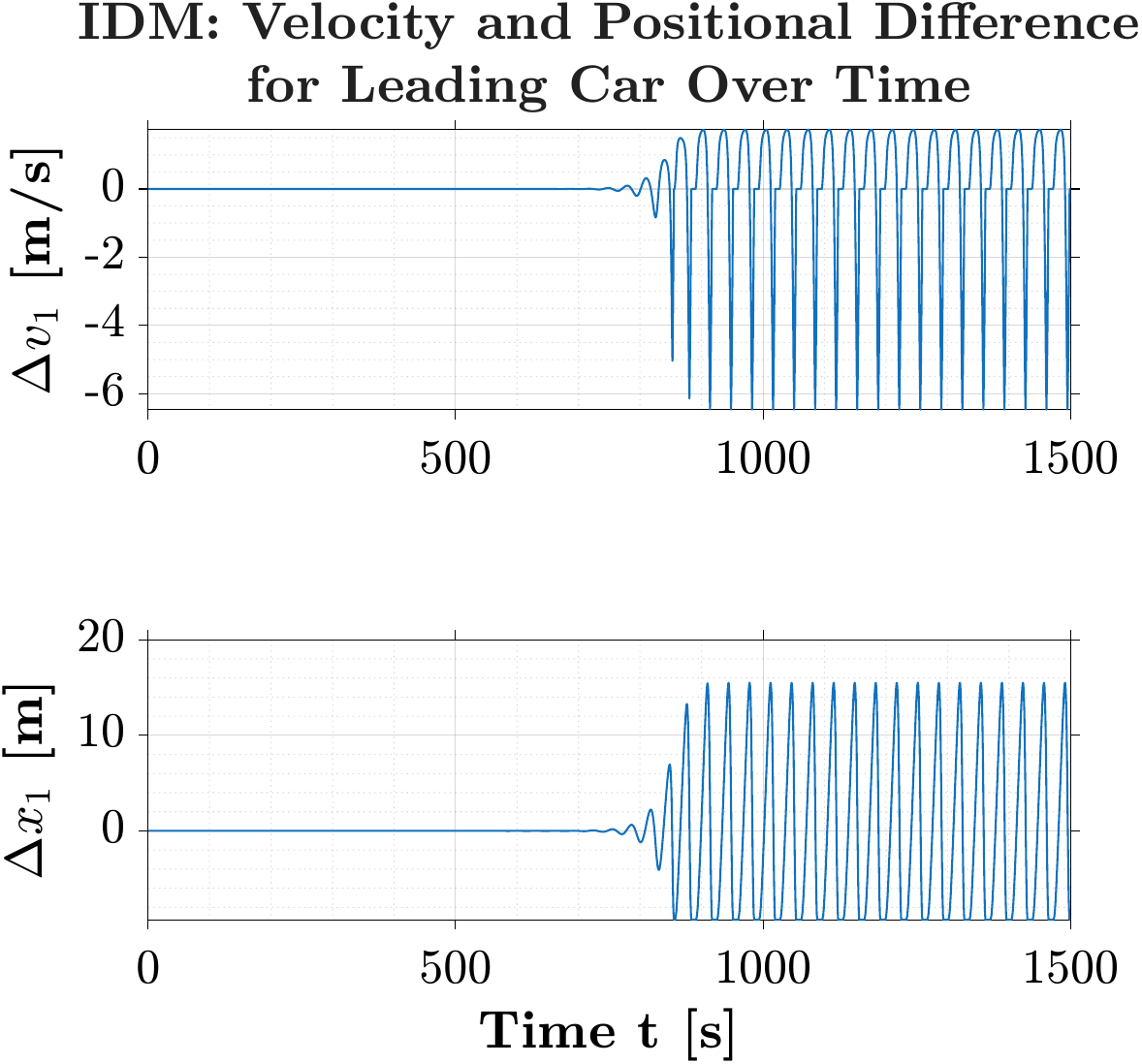}
    \includegraphics[width=0.49\linewidth]{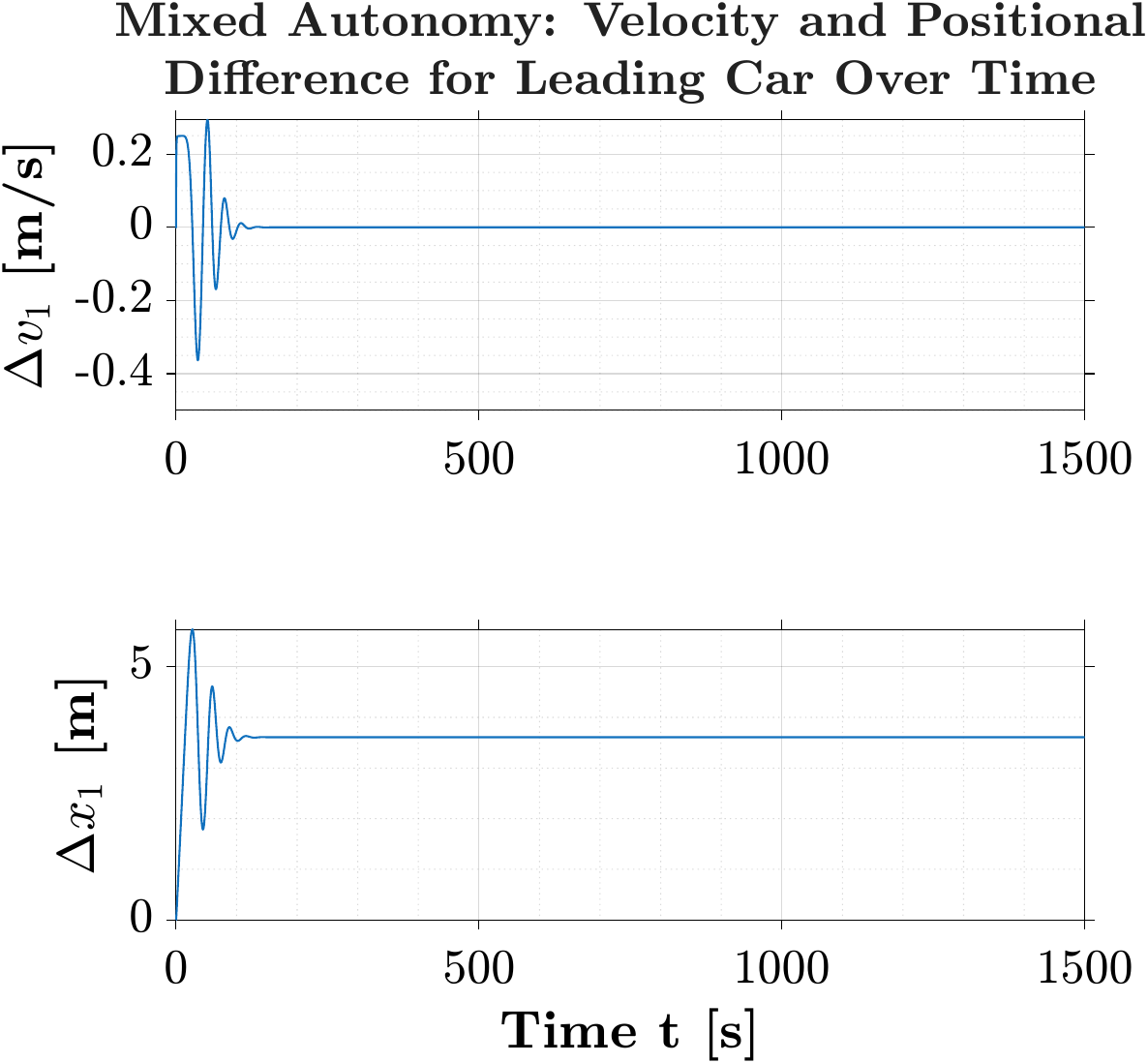}
    \caption{\textbf{Left:} Plots showing the difference between the velocity or position of vehicle 1 and its leader in the IDM simulation; \textbf{Right:} IDM-Followerstopper mixed autonomy simulation. The positional difference was subtracted by unity spacing, $10m$, to emphasize deviation from the expected spacing.}
    \label{fig:IDM_vs_gap}
        \vspace{-16pt}
\end{figure}

\begin{figure}[H]
    \centering
    \includegraphics[width=0.49\linewidth]{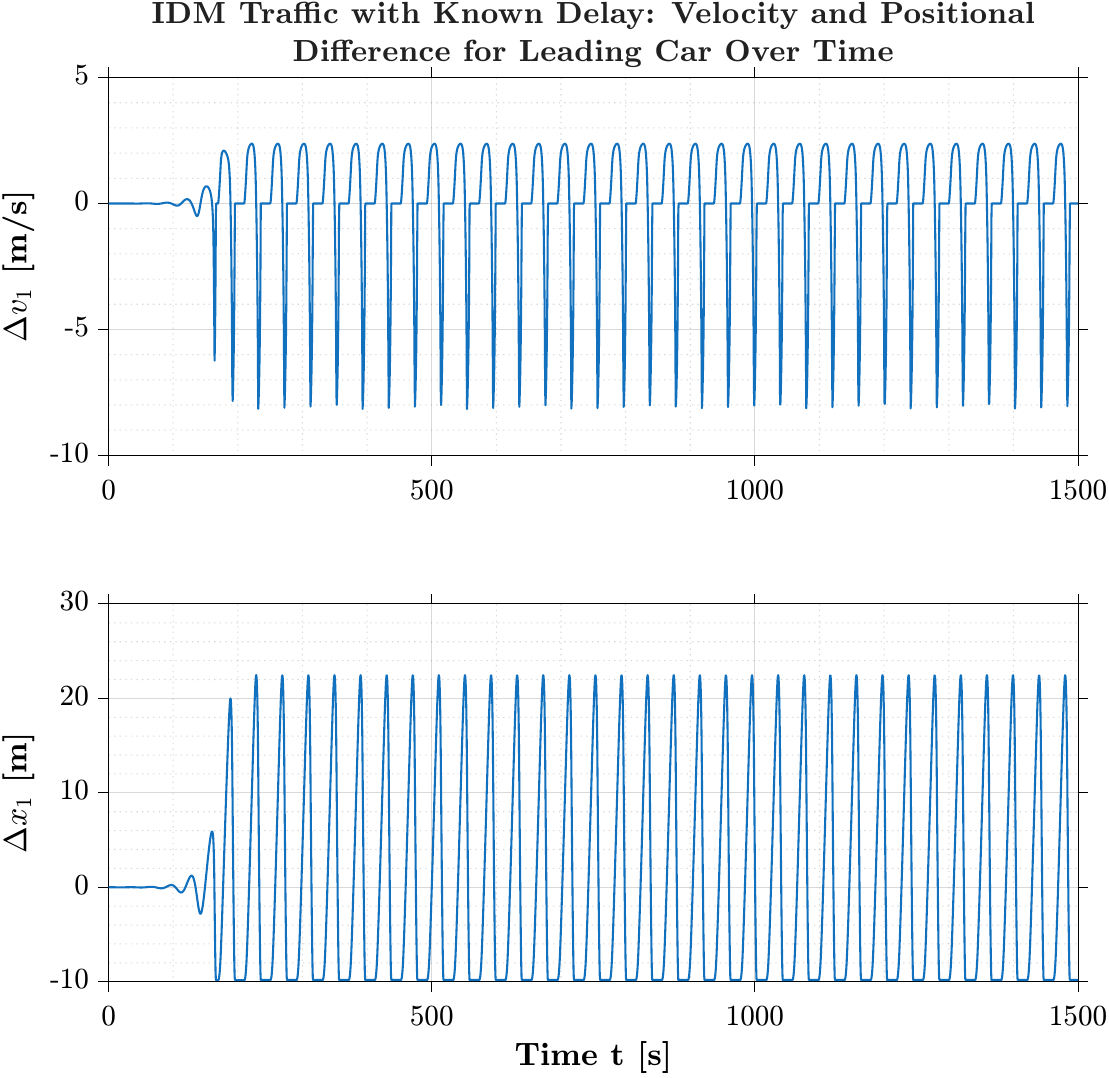}
    \includegraphics[width=0.49\linewidth]{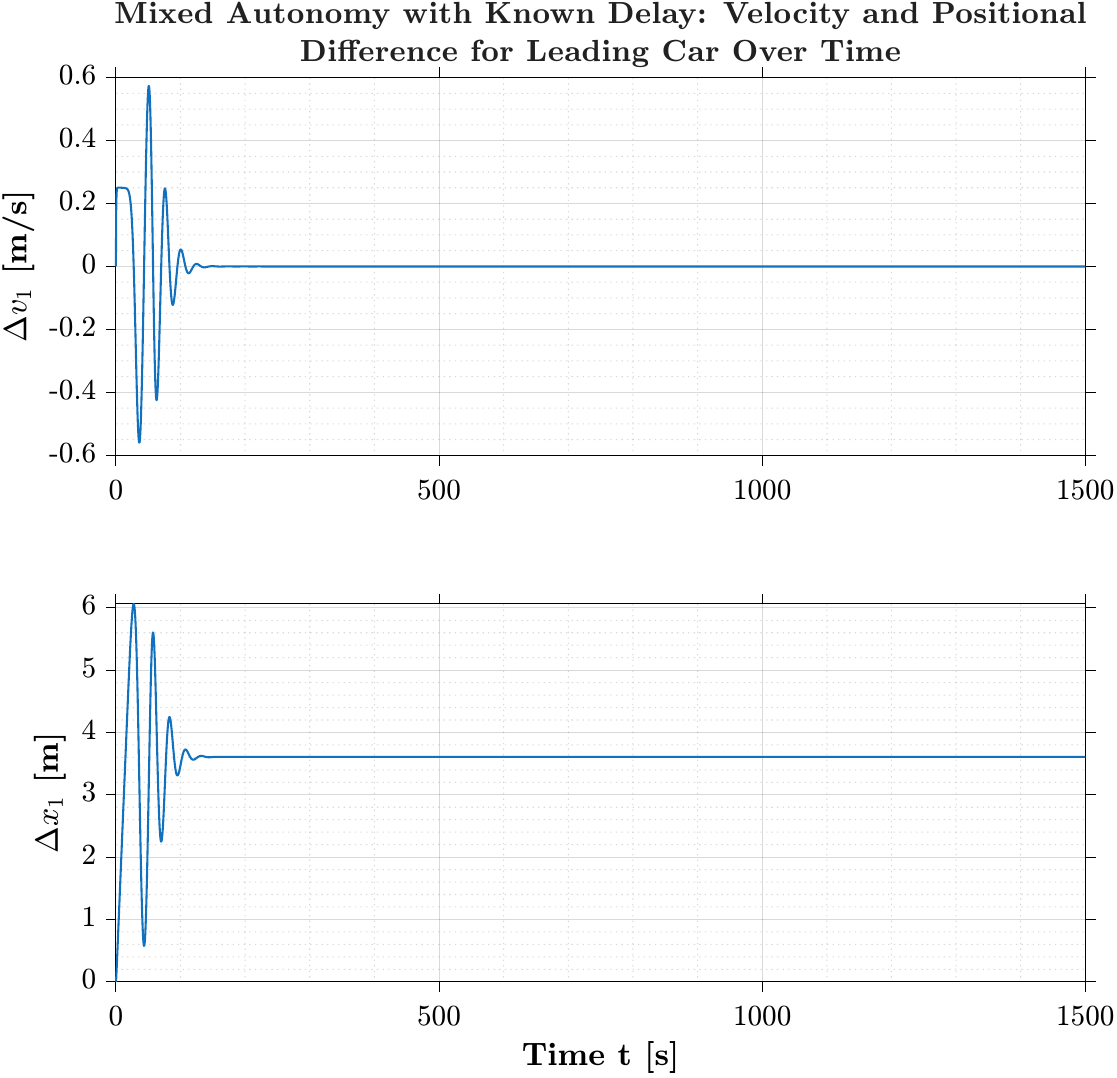}
    \caption{\textbf{Left:} plots showing the difference between the velocity or position of vehicle 1 and its leader in the delayed IDM simulation; \textbf{Right:} delayed IDM with Followerstopper enabled mixed-autonomy simulation. The positional difference was subtracted by unity spacing, $10m$, to emphasize deviation from expected spacing.}
    \label{fig:D_IDM_vs_gap}
    \vspace{-16pt}
\end{figure}

The Velocity and space difference plots demonstrate the velocity profile of the stop-and-go waves. When looking at the mixed-autonomy scenario, it is clear that Followerstopper maintained a larger space headway than the expected $10m$. This allowed Followerstopper to dissipate stop-and-go waves, without propagating them to its follower vehicle. The velocity difference and spacing differences shown in Figure~\ref{fig:D_IDM_vs_gap} indicate a similar response.

\subsection{Phase Space Diagram}

\begin{figure}[H]
    \centering
    \includegraphics[width=0.49\linewidth]{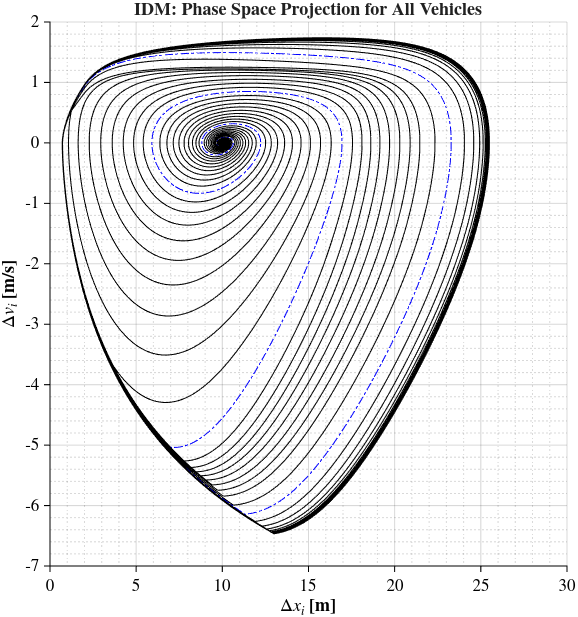}
    \includegraphics[width=0.49\linewidth]{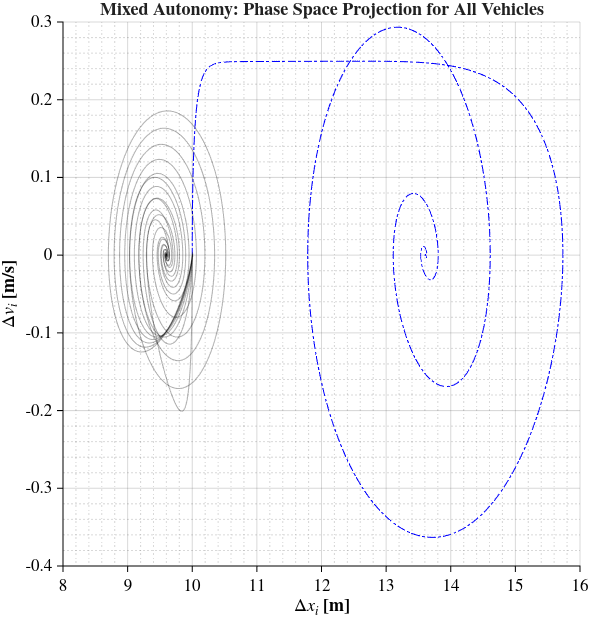}
    \caption{\textbf{Left:} Phase space projection diagram showing the system dynamics in the IDM simulation; \textbf{Right:} Phase space projection diagram showing the system dynamics in the mixed autonomy simulation. The dotted-blue line shows the first vehicle in the platoon, which corresponds to Followerstopper in Simulation 3.}
    \label{fig:IDM_PS}
        \vspace{-16pt}
\end{figure}

In Figure~\ref{fig:IDM_PS} (left), we can see that the dynamics of the system suggest a stable limit cycle rotating around an unstable focus. This is not ideal for traffic flow as it can lead to continuous formation of stop-and-go waves. In contrast, Followerstopper transformed the unstable focus into a stable focus, attracting velocities in a local neighborhood toward the critical point that corresponds to the uniform flow manifold. 

\begin{figure}[H]
    \centering
    \includegraphics[width=0.49\linewidth]{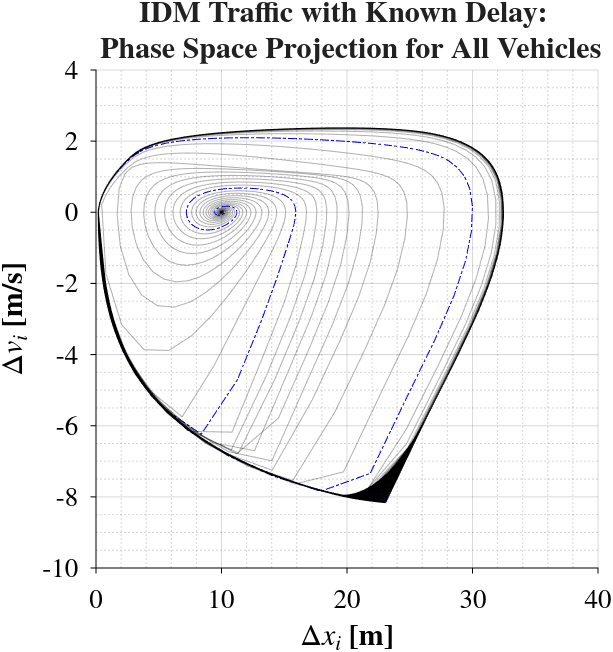}
    \includegraphics[width=0.49\linewidth]{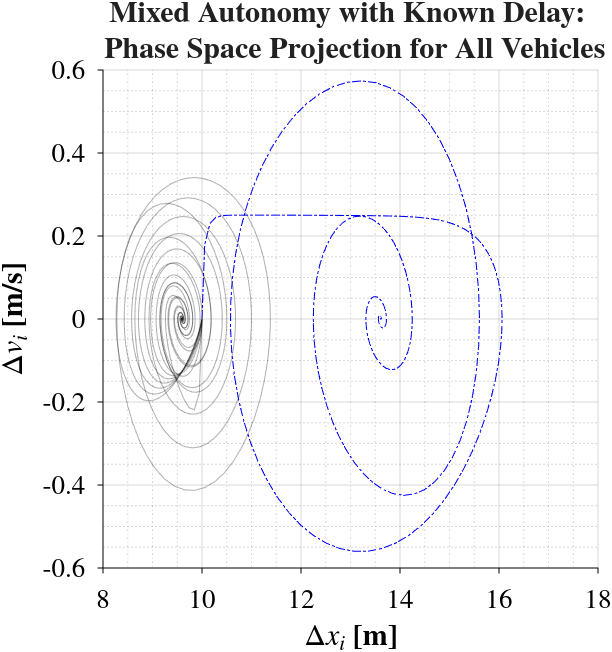}
    \caption{\textbf{Left:} Phase space diagram showing the system dynamics in the delayed IDM simulation; \textbf{Right:} Phase space diagram showing the system dynamics in the delayed IDM with Followerstopper enabled mixed autonomy simulation. The dotted-blue line shows the first vehicle in the platoon, which corresponds to Followerstopper in Simulation 4.}
    \label{fig:D_IDM_PS}
\end{figure}
From the Phase space diagram, Figure~\ref{fig:D_IDM_PS} (left), we can see that the fundamental dynamics of the delayed system are not radically different from the dynamics of the ideal IDM in simulation 1, Figure~\ref{fig:IDM_PS} (left). However, the limit cycle is wider, indicating a greater difference between the maximum and minimum gaps and the maximum and minimum velocities. The results in Followerstopper simulation are similar, showing a wider velocity and gap difference, but similar dynamics, dissipating the stop-and-go waves before they  fully emerge.

\subsection{Lyapunov Exponents for All Simulations}

\begin{table}[H]
    \centering
    \renewcommand{\arraystretch}{1} 
    \begin{tabular}{|c||c|}  
        \hline
        \textbf{Simulation} & \textbf{Maximal Lyapunov Exponent} \\
        \hline
        Ideal IDM & 0.0780 \\
        \hline
        Ideal IDM with FS & -0.0741 \\
        \hline
        Delayed IDM & 2.8572 \\
        \hline
        Delayed IDM with FS & -0.0656 \\
        \hline
    \end{tabular}
    \caption{Maximal Lyapunov exponent for each simulation. A negative exponent indicates stability, while a positive exponent indicates instability.}
    \label{tab:lyapunov_matrix_ch3}
    \vspace{-5pt}
\end{table}

The Lyapunov exponents in Table~\ref{tab:lyapunov_matrix_ch3} confirm the system's stability characteristics across different scenarios. The introduction of a $0.5$-second delay drastically reduces stability, as shown by the high positive exponent for the delayed IDM simulation; this aligns with the early emergence of stop-and-go waves observed in that case. Conversely, Followerstopper controller successfully induces stability in both the ideal and delayed environments, yielding negative Lyapunov exponents. The
Lyapunov exponent for the delayed IDM with Followerstopper simulation is less negative than that of the Lyapunov exponent
for the ideal IDM with Followerstopper simulation, which is
consistent with expectations.
	
	\section{Summary and Conclusions}
	\label{sec:conclusion}
	   This paper examined the dynamical system constructed from simulating the trajectories of IDM vehicles on a ring road under different conditions, including time delay and a mixed autonomy simulation with  Followerstopper model. From these simulations, it is clear that the uniform flow manifold for the ideal and delayed IDM ring road simulations is unstable from a dynamical systems perspective. This can be observed from the maximum Lyapunov exponents and phase space plots, showing the vehicle velocity difference and positional gap spiraling around the unstable uniform flow manifold, before settling within a limit cycle. This behavior indicates a stop-and-go wave affecting the system. The cause for these stop-and-go waves is the propagation and magnification of small deviations in position and velocity. These deviations magnify, leading to the vehicle desiring a negative velocity to maintain its desired space headways. We enforce the non-negative velocity assumption in the simulation, as on highways, cars don't move backward. As a result, the velocity of the cars stays at $0m/s$ until the traffic in front of it de-congests. In contrast, the mixed autonomy simulations with  Followerstopper showed stability from a dynamical systems perspective. This is clear from the maximal Lyapunov exponent as well as the phase plots. Followerstopper maintained a variable, real-time updating space headway, where different thresholds induce different behavior. This allowed Followerstopper to dissipate these small perturbations before they could fully magnify into stop-and-go waves. Further research should focus on more complex vehicle simulations with more cars or different speed profiles. Alternatively,  we may look to extend the scenario to consider multiple lanes of traffic, with frequent lane changes.

	\section*{ACKNOWLEDGMENTS}

    {
    \footnotesize
	\bibliographystyle{IEEEtran}
	\bibliography{root} 
	}
\end{document}